\documentclass[aps,prd,twocolumn,groupedaddress]{revtex4}
\usepackage{amsmath, amsfonts, graphicx}
\usepackage{epstopdf}

\begin{document}

\title{Higgs Masses in the Four Generation MSSM}

\author{Sean Litsey}\email[]{sean.litsey@yale.edu}
\affiliation{Department of Physics, Yale University, New Haven, CT 
06511-8499 USA}

\author{Marc Sher}\email[]{mtsher@wm.edu}
\affiliation{Particle Theory Group, Department of Physics, College of William and Mary, Williamsburg, VA 23187, USA}

\date{\today}

\begin{abstract}
    In the Minimal Supersymmetric Standard Model (MSSM) with three 
    generations of fermions, there is a stringent upper bound on the 
    mass of the lightest neutral Higgs $h$, and the mass of the 
    charged Higgs $H^{+}$,
    must be close (within tens of GeV) to the heavier neutral Higgs, 
    $H$, 
    and the pseudoscalar Higgs $A$.  In this Brief Report, we show that in the four generation 
    MSSM, the upper bound on the $h$ mass
    is much higher, as high as 400 GeV, and $H^{+}$ is 
    generally much heavier than the $A$, allowing the  
    $H^{+}\rightarrow A W^+$ decay, potentially changing search strategies for both the 
    charged Higgs and the pseudoscalar.   The $H$ mass, on the other hand, remains within 
    tens of GeV of the charged Higgs mass.
\end{abstract}

\pacs{}

\maketitle


\section{Introduction}

One of the most important constraints on supersymmetric models comes 
from considerations of the mass of the lightest neutral Higgs, $h$.
In the minimal supersymmetric standard model (MSSM), the mass of the $h$ 
field at tree level \cite{inoue,hhg} must be less than the $Z$ mass, but radiative 
corrections (primarily due to top quark/squark loops) can lift this 
to $135$ GeV \cite{haber}.   While non-minimal supersymmetric 
models can evade this bound somewhat, it is very difficult \cite{kane} to raise 
the upper bound by more than a few tens of GeV.

Another constraint comes from the mass of the charged Higgs boson, 
$H^{+}$.   At tree level,  one has \cite{inoue,hhg}
\begin{equation}
    M_{H^{+}}^{2} = M_{A}^{2} + M_{W}^{2}
    \end{equation}
    where $A$ is the pseudoscalar Higgs mass.   Thus, the charged 
    Higgs mass cannot be much heavier than the pseudoscalar, and the 
    decay $H^{+}\rightarrow A W^+$ is kinematically forbidden\cite{latest}.  Similar 
    considerations involving the mass of the heavier neutral scalar, 
    $H$, show that at tree level 
\begin{equation}
    M_{h}^{2}+M_{H}^{2} = M_{H^{+}}^{2} + M_{Z}^{2}-M_{W}^{2}
    \end{equation}
which also shows that the charged Higgs cannot decay into $H W^+$.
Radiative corrections to these formulae have been calculated 
\cite{diaz} and, as 
in the neutral case, are typically tens of GeV.

In this Brief Report, we note that one of the simplest extensions of 
the MSSM substantially changes these bounds; the addition of a 
sequential fourth generation.   Interest in a fourth chiral 
generation (with a sufficiently heavy neutrino to avoid contributing 
to the Z width) has fluctuated over the years \cite{fhs,holdom}.
In the early part of 
this decade, it was believed by many that electroweak radiative 
corrections ruled out a chiral fourth generation \cite{pdg}.  However, this 
belief was based on the assumption that the quark masses were 
degenerate.  It has been shown \cite{subsequent} that relaxing this assumption allows a 
fourth generation, and for a mass splitting of $O(10)\% $, the S and 
T corrections  fit inside the one-sigma 
error ellipse \cite{kpst}.   

Since corrections to the above mass relations vary  as the fourth 
power of the top quark mass, it is clear that they could be 
substantially enhanced by a fourth generation.  Since the mass 
splitting between the $b'$ and $t'$ quarks is much smaller than their 
masses, we will assume that they are degenerate in our analysis.  We will consider 
masses between $250$ and $400$ GeV.  The lower bound is somewhat 
lower than 
current Tevatron limits \cite{current}, since it might be possible to weaken these limits  
if supersymmetric decay modes are allowed.  The upper bound comes from 
the requirement that perturbation theory be valid.  Generically, the 
requirement that the Yukawa couplings, $g_{Y}$ in the MSSM satisfy 
the condition $g_{Y}^{2}/4\pi < 1$ implies \cite{dms} that  $1/x < 
\tan\beta < x$, where $x=
  \sqrt{2\pi (v/M)^{2}-1} $,
 $\tan\beta$ is the ratio of vacuum expectation values, $M$ 
is the fermion mass, and $v=246$ GeV.
This window of $\tan\beta$ closes for masses greater than $500$ GeV.  
Since the Yukawa coupling increases with scale, and we want the model 
to be valid up to at least a TeV, we will set the upper bound to 
$400$ GeV.

In the next section, we look at the lightest neutral scalar, and then 
consider the mass relations involving the charged Higgs scalar in 
Section III.  Section IV contains our conclusions.

\section{The lightest neutral Higgs mass}

As noted above, $\tan\beta$ must be fairly close to one in order for
perturbation theory be valid.   In this case, the tree level value of 
the lightest neutral Higgs mass is negligible, and the entire mass 
must come from radiative corrections.  In the three generation case, 
these corrections have been calculated, first in Ref. \cite{haber}, 
and then with increasing precision by many others \cite{others}.  
Generally, the contributions from radiative corrections 
will not be sufficient to increase the mass above $100$ GeV (if the 
tree level mass is negligible).

In the case of four generations, this is not the case.  Some early 
works exploring four generation supersymmetry models\cite{early} did 
not really address the possibility of fourth generation quarks above 
300 GeV.  More recently, an exploration of the electroweak phase 
transition in the four generation MSSM appeared\cite{recent}, but 
it used a value of $\tan{\beta}$ which put the $b'$ Yukawa coupling 
far outside the region of perturbative validity.  Last year, Murdock, 
et al.\cite{nandi} discussed ways of extending the validity of 
perturbation theory in the four generation MSSM.   The only recent 
discussion of the lightest Higgs mass in the four generation MSSM was 
the work of Fok and Kribs \cite{fok}, who plotted the lightest Higgs 
mass as a function of the heavy quark mass.  They did restrict 
the model to the case in which the scalar $t'$ and $b'$ masses were within 
twenty percent of the $t'$ and $b'$ masses, since that was necessary 
to obtain sufficient electroweak baryogenesis.  

The expression used by Fok and Kribs \cite{fok} in the limit in which 
mixing between scalar quarks is ignored, all squarks have the same 
mass, the heavier Higgs bosons are all very heavy and $\tan\beta=1$ is
\begin{equation}
    m^{2}_{h}=\sum_{f=t,t',b'} {3\over 2\pi^{2}}{m_{f}^{4}\over 
    v^{2}}\ln{{m_{SUSY}^{2}\over m_{f}^{2}}}
    \end{equation}
where $v=246$ GeV.  A complete expression which includes two-loop 
leading logs, as well as squark mixing, can be found in the review 
article of Carena and Haber \cite{carenahaber}.   We have used that 
complete expression, and have plotted the results in Figure 1 for 
various values of the SUSY scale.

 \begin{figure}
        \includegraphics[scale=.55]{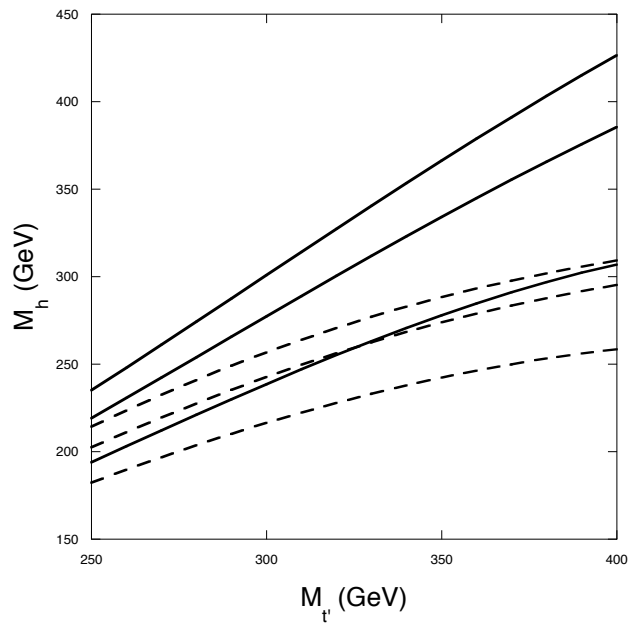}
        \caption{Mass of the lightest Higgs in the MSSM as a function 
	of the $t'$ mass.  The squark masses are presumed to be equal 
	and with a value of $M_{SUSY}$  The solid (dashed) lines are for a 
	pseudoscalar mass of $1000$ ($500$) GeV, and the curves, from 
	the lower curve up, correspond to $M_{SUSY}= 600, 800, 1000$ 
	GeV, respectively }
	\end{figure}

In the graph shown, we have taken the mass of the pseudoscalar to be either $500$ or $1000$ GeV, and have fixed 
$\tan\beta=1$.  The results are very insensitive to squark 
mixing and the precise values of $\tan\beta$ and $\mu$.   However, 
they are  sensitive to the pseudoscalar mass.  For a pseudoscalar 
mass of several TeV, the results agree with that given in Eq. (3), 
but for pseudoscalar masses of $1000$ ($500$) GeV, the results 
(for $m_{t'}=400$ GeV) are smaller than that of Eq. (3) by $15$ ($35$) percent.

We see that a large mass for the lightest Higgs boson in the 
four generation MSSM not only possible, but is likely for SUSY scales above several 
hundred GeV.   Even a mass of close to $400$ GeV cannot be excluded.

\section{The charged Higgs mass}

As was noted earlier,  Eq. (1) prevents the decay $H^{+}\rightarrow
A W^+$ at tree level.    This makes detection of the charged Higgs 
quite difficult, since the $H^{+}\rightarrow \bar{t} b, \bar{b}t$ decay has very 
large backgrounds.  Radiative corrections to Eq. (1) can be
found in Ref. \cite{diaz}, and when these are modified to account for a
fourth generation, the decay may no longer be kinematically forbidden.
One obtains an expression\cite{diaz} for $\Delta m \equiv m_{H^+} - m_A$
 which depends primarily on $m_{t'}$
and $m_{SUSY}$.  The $\tan\beta$ dependence is very small due to the 
above requirement of the validity of perturbation theory.
The dependence on $m_{t'}$ is
illustrated in Figure 2 for several values of $m_{SUSY}$ and
$\tan\beta$ fixed to 1.     We see that the $H^{+}\rightarrow
A W^+$ decay becomes kinematically allowed for much of parameter space.

 \begin{figure}
        \includegraphics[scale=.55]{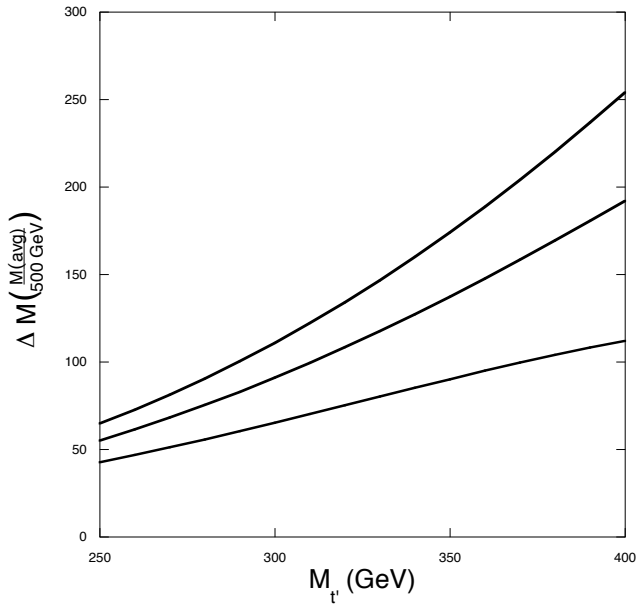}
        \caption{The mass difference between the charged Higgs scalar 
	and the pseudoscalar in the MSSM as a function 
	of the $t'$ mass.  The squark masses are presumed to be equal 
	and with a value of $M_{SUSY}$  The curves, from 
	the lower curve up, correspond to $M_{SUSY}= 600, 800, 1000$ 
	GeV, respectively.  $M(avg)$ is the average of the charged scalar and pseudoscalar masses. }
	\end{figure}

The situation is different with respect to the decay
$H^{+}\rightarrow H W^+$.  Radiative corrections to  Eq. (2) are given 
in Ref. \cite{carenahaber}, and these can easily be generalized to a 
fourth generation.  We
find that $m_H$ is never less than $m_{H^+}-m_W$, and in fact usually
$m_H > m_{H^+}$ by tens of GeV.  Therefore this decay process is
still forbidden, even with higher order corrections.

In addition to the possibility of the decay process $H^{+}\rightarrow 
A W^+$, one has the conventional processes 
$H^{+}\rightarrow \bar{t} b. \bar{b}t$ and, for sufficiently heavy charged Higgs bosons, 
  $H^{+}\rightarrow 
\bar{t'}b', \bar{b'}t'$  Furthermore, one can also have 
$H^{+}\rightarrow h W^+$.  The latter decay will be small in this 
case.  The reason is that in the limit of $M_{A} >> 100$ GeV, one 
has\cite{hhg} $\alpha=-\beta$, and thus the coupling of the light 
Higgs to the charged Higgs and a $W$, which is proportional to 
$\cos(\alpha-\beta)$, vanishes as $\tan\beta\approx 1$.    Of course, one might also have various supersymmetric decay modes.   We will only consider decays into $\bar{t}b$, $\bar{t'}b'$ and $A W^+$.

   \begin{figure}
        \includegraphics[scale=.55]{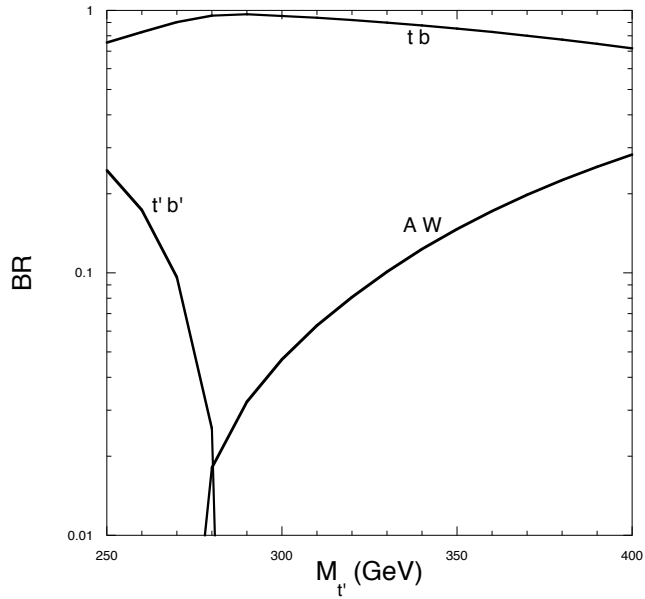}
        \caption{The branching ratio for $H^{+}\rightarrow A W^+$,  $H^{+}\rightarrow \bar{t} b, \bar{b}t$ and $H^{+}\rightarrow 
\bar{t'}b', \bar{b'}t'$ versus $m_{t'}$, choosing $M_{SUSY}=1000$ GeV and $M_A=500$ GeV. }
	\end{figure}

In Figure 3, 
we plot the branching ratios
of $H^{+}\rightarrow A W^+$,  $H^{+}\rightarrow \bar{t}  b, \bar{b}t$ and $H^{+}\rightarrow 
\bar{t'} b', \bar{b'} t'$ versus $m_{t'}$, choosing $M_{SUSY}=1000$ GeV and $m_A=500$ GeV.    Note that if the pseudoscalar mass is $1000$ GeV, then one can see from Figure 2 that $H^{+}\rightarrow A W^+$ is only allowed for very large values of $m_{t'}$, and the branching ratio for this mode is always less than a percent.
For $m_A = 500$ GeV,  the $H^{+}\rightarrow 
\bar{t'} b', \bar{b'} t'$ decay rapidly becomes kinematically inaccessible, and the branching ratio for $H^{+}\rightarrow A W^+$ can be as large as $20\%$.
Note that in this case, the pseudoscalar will primarily decay into  $\bar{t} t$, 
and observation of the associated $W$ may be the only 
way to detect it\cite{last}.

\section{Conclusions}

In the MSSM, the Higgs sector is very tightly restricted.  There is 
an upper bound of approximately $130$ GeV on the lightest Higgs mass 
and the charged Higgs mass cannot differ by more than tens of GeV 
from the pseudoscalar and heavier neutral Higgs masses, thus ruling 
out two-body decays into these states.   In this Brief Report, 
motivated by recent interest in four generation models, we have 
considered the effects of radiative correction in the four generation 
MSSM.  The upper bound on the lightest Higgs mass is substantially 
increased, and can be as large as $400$ GeV for reasonable parameter 
values.   The mass splitting between the charged Higgs and 
pseudoscalar is increased to the point where the $H^{+}\rightarrow 
A W^+$ decay is kinematically allowed, whereas the splitting between 
the charged Higgs and heavy neutral scalar remains small. 

Should a fourth generation exist, it will likely be discovered shortly at the LHC.   Our main point is that radiative corrections from these quarks can have major implications for the Higgs sector, not only for the MSSM, but other extensions of the Standard Model.

\begin{acknowledgments}
    We are grateful to Chris Carone, David Reiss and Graham C. Kribs for
    useful discussions.  This work was supported by the National Science
    Foundation grant NSF-PHY-0757481 and the REU grant NSF-PHY-0755262.
\end{acknowledgments}

\end{document}